\documentclass[twocolumn]{aastex62}

\usepackage{amsmath}
\usepackage{amssymb}
\usepackage{empheq}
\usepackage{mathrsfs}
				
\usepackage{amssymb}

\begin{document}

\title{Injection of Inner Oort Cloud Objects Into the Distant Kuiper Belt by Planet Nine}

\author{Konstantin Batygin}
\affiliation{Division of Geological and Planetary Sciences California Institute of Technology, Pasadena, CA 91125, USA}

\author{Michael E. Brown}
\affiliation{Division of Geological and Planetary Sciences California Institute of Technology, Pasadena, CA 91125, USA}

\begin{abstract}
The outer solar system exhibits an anomalous pattern of orbital clustering, characterized by an approximate alignment of the apsidal lines and angular momentum vectors of distant, long-term stable Kuiper belt objects. One explanation for this dynamical confinement is the existence of a yet-undetected planetary-mass object, “Planet Nine (P9)’’. Previous work has shown that trans-Neptunian objects, which originate within the scattered disk population of the Kuiper belt, can be corralled into orbital alignment by Planet Nine's gravity over $\sim$Gyr timescales, and characteristic P9 parameters have been derived by matching the properties of a synthetic Kuiper belt generated within numerical simulations to the available observational data. In this work, we show that an additional dynamical process is in play within the framework of the Planet Nine hypothesis, and demonstrate that P9-induced dynamical evolution facilitates orbital variations within the otherwise dynamically frozen inner Oort cloud. As a result of this evolution, inner Oort cloud bodies can acquire orbits characteristic of the distant scattered disk, implying that if Planet Nine exists, the observed census of long-period trans-Neptunian objects is comprised of a mixture of Oort cloud and Kuiper belt objects. Our simulations further show that although inward-injected inner Oort cloud objects exhibit P9-driven orbital confinement, the degree of clustering is weaker than that of objects originating within the Kuiper belt. Cumulatively, our results suggest that a more eccentric Planet Nine is likely necessary to explain the data than previously thought. 
\end{abstract}

\keywords{planets and satellites: dynamical evolution and stability}

\section{Introduction}

Over the course of the past three decades, the dynamical architecture of the solar system’s trans-Neptunian region has steadily come into sharper focus. In turn, the unprecedented level of detail unveiled by observational surveys has repeatedly challenged our understanding of the solar system’s markup, as well as the hitherto standard model of its long-term evolution \citep{2020CeMDA.132...12S}. The resulting paradigm shift towards a instability-driven model of the solar system’s early dynamics has resolved much of the tension between theory and observations, and today, the physical mechanisms that shaped the structure of the proximate ($a\lesssim50\,$AU) Kuiper belt have been broadly delineated (\citealt{2020tnss.book...25M} and the references therein). In contrast, a full understanding of the anomalous orbital architecture exhibited by distant trans-Neptunian objects -- and its dynamical origins -- remains elusive.

\begin{figure*}[tbp]
\centering
\includegraphics[width=0.8\textwidth]{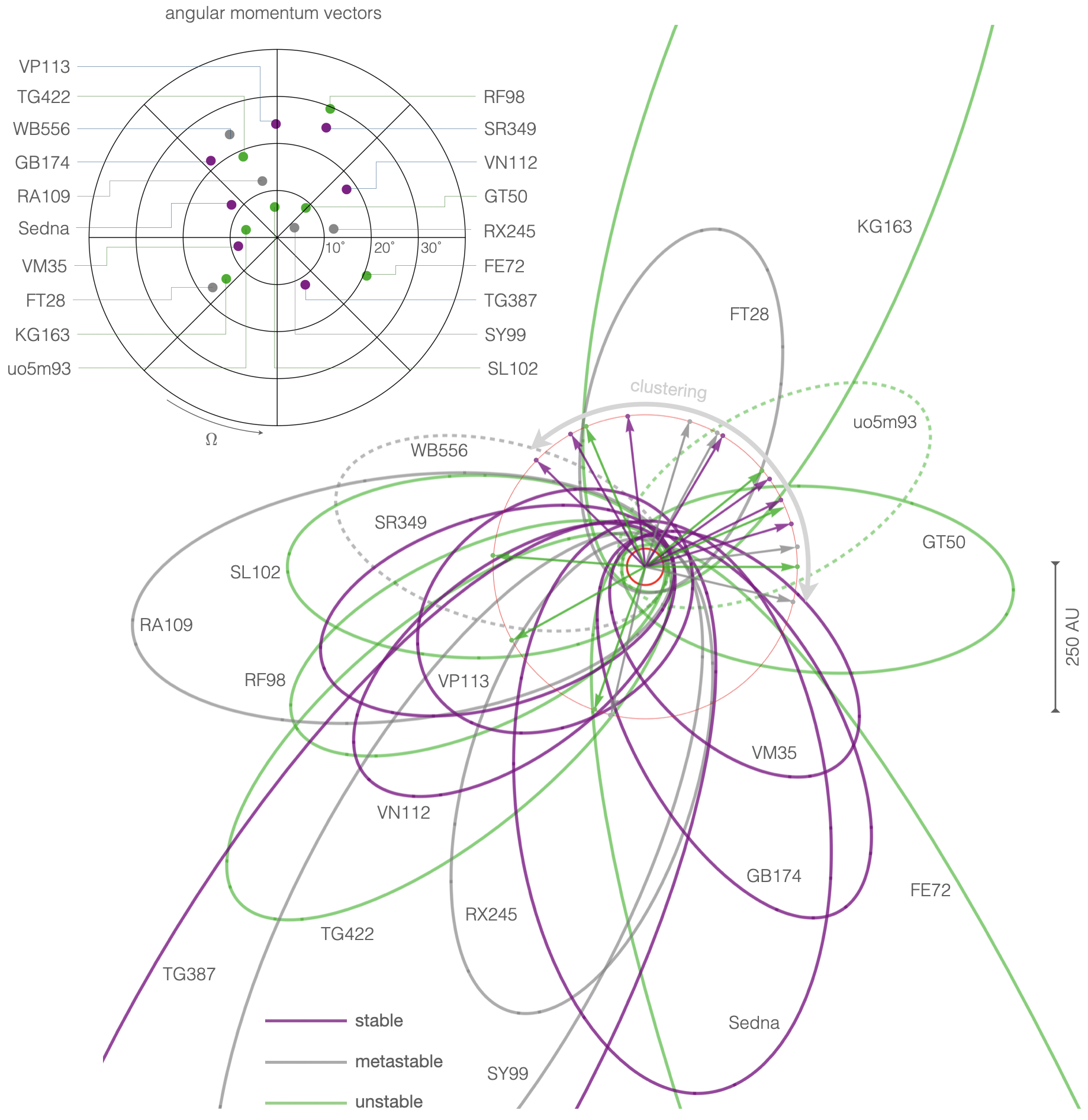}
\caption{
Census of trans-Neptunian objects with $a>250\,$AU, $q>30\,$AU, and $i<40\,\deg$. The orbits of objects that experience rapid orbital diffusion are shown in green, while the orbits of stable and slowly diffusing KBOs are depicted in purple and gray, respectively. The dotted orbits correspond to KBOs whose parameters and uncertainties are not reported in the JPL Small Body Database, and whose stability was estimated from a single $4\,$Gyr integration of the nominal orbit rather than similar integrations of ten clones of each object. The $i-\Omega$ polar inset informs the tilt of the orbital angular momentum vectors. Stable and metastable objects have apsidal lines that cluster around $\langle\varpi \rangle\sim60\,\deg$, and form a mean plane that is inclined with respect to the ecliptic by $\langle i \rangle\sim10\deg$, with a mean longitude of ascending node of $\langle\Omega \rangle\sim90\deg$.
}
\label{F:orbits}
\end{figure*}

Perhaps the most striking peculiarity displayed by the extended Kuiper belt is the approximate alignment of long-period ($P\gtrsim4{,}000\,$yr) orbits in physical space\footnote{Simultaneous alignment of orbital planes and eccentricity vectors requires the orbits to have similar longitudes of perihelion, $\varpi$, inclinations, $i$, and longitudes of ascending node, $\Omega$. Importantly, we note that the grouping of the arguments of perihelion, $\omega=\varpi-\Omega$, was first pointed out by \citet{2014Natur.507..471T}.} \citep{BB2016}. That is, trans-Neptunian objects with semi-major axes in excess of 250 AU and inclinations smaller than 40 deg have apsidal lines and angular momentum vectors that cluster together in an unexpected manner (Figure \ref{F:orbits}). Although this clustering is visible by eye, it is important to note the degree of orbital confinement is not uniform, and instead depends sensitively on dynamical stability. In particular, objects that experience comparatively rapid orbital diffusion due to periodic interactions with Neptune (shown as green ellipses on Figure \ref{F:orbits}) are much less tightly confined than their stable and metastable counterparts\footnote{For the purposes of this work, we define the tiers of orbital stability as follows: if more than $20\%$ of the clones of a given object eject over the course of a 4 Gyr integration, an object is deemed unstable. If more than $20\%$ of the clones of a given object change their semi-major axes by a factor of two during a 4 Gyr integration but do not eject, an object is deemed metastable. If neither of these events occurs, we deem such an object stable.} (shown as purple and gray ellipses on Figure \ref{F:orbits}, respectively).

A number of explanations have been proposed for the origins of the alignment depicted in Figure (\ref{F:orbits}). Arguably the most rudimentary of them is the notion that the alignment is not real, and is instead a figment of observational bias or statistical chance. To this end, individual surveys -- which have only searched limited areas of the sky -- have been unable to overcome their own observational biases sufficiently to rigorously determine the absence or presence of orbital alignment \citep{2017AJ....154...50S,2020PSJ.....1...28B,2021arXiv210205601N}. Nevertheless, a combined observability analysis of all available data shows that, after taking observational biases into account, distant KBOs are jointly clustered in eccentricity and angular momentum vectors at the $\sim99.8\%$ significance level \citep{2017AJ....154...65B, BB2019}. Thus, if we interpret the picture insinuated by the data at face value, the observed clustering requires a sustained gravitational torque that maintains orbital alignment against differential precession induced by Jupiter, Saturn, Uranus and Neptune. 


In principle, such a torque could come from any distant massive object, including a lopsided massive disk of icy material \citep{2018AJ....156..141M,2019AJ....157...59S}, a primordial black hole \citep{2020PhRvL.125e1103S}, an asymmetric dark matter halo, etc. In this work, we opt for what we consider a somewhat less exotic alternative, and envision that the alignment of distant orbits is sustained by a yet-unseen planet, ``Planet Nine", occupying a mildly eccentric, slightly inclined orbit with a period of order $\sim10{,}000\,\,$years \citep{BB2016,2016ApJ...824L..23B}. We note however, that because the dominant mode of P9-induced dynamics is secular (i.e., stemming from phase-averaged gravitational potential of P9), our results are not sensitive to the composition of P9 -- only its mass and its orbit.

\section{Hypothesis}

In a recent effort \citep{BABB}, we have carried out an extensive suite of $N-$body simulations of the solar system’s long-term evolution in presence of Planet Nine, and have demonstrated that a synthetic Kuiper belt sculpted by a $m_9=5\,M_{\oplus}$ P9 residing on a $a_9=500\,$AU, $e_9=0.25$, $i_9=20\,\deg$ orbit adequately matches the observations. A subtle limitation of these calculations, however, is that they treat the solar system as an isolated object, ignoring the effects of passing stars, both during the solar system’s infancy and during its long-term field evolution, as well as the gravitational tides of the birth cluster and the Galaxy. As such, all of the material that comprises the distant ($a\gtrsim250\,$AU) Kuiper belt within the framework of these numerical experiments is envisioned to have been sourced from inside of 30AU, and emplaced into the trans-Neptunian region by scattering off of the giant planets during the solar system’s early transient instability \citep{2008Icar..196..258L,2018ARA&A..56..137N}. 

\begin{figure*}[tbp]
\centering
\includegraphics[width=\textwidth]{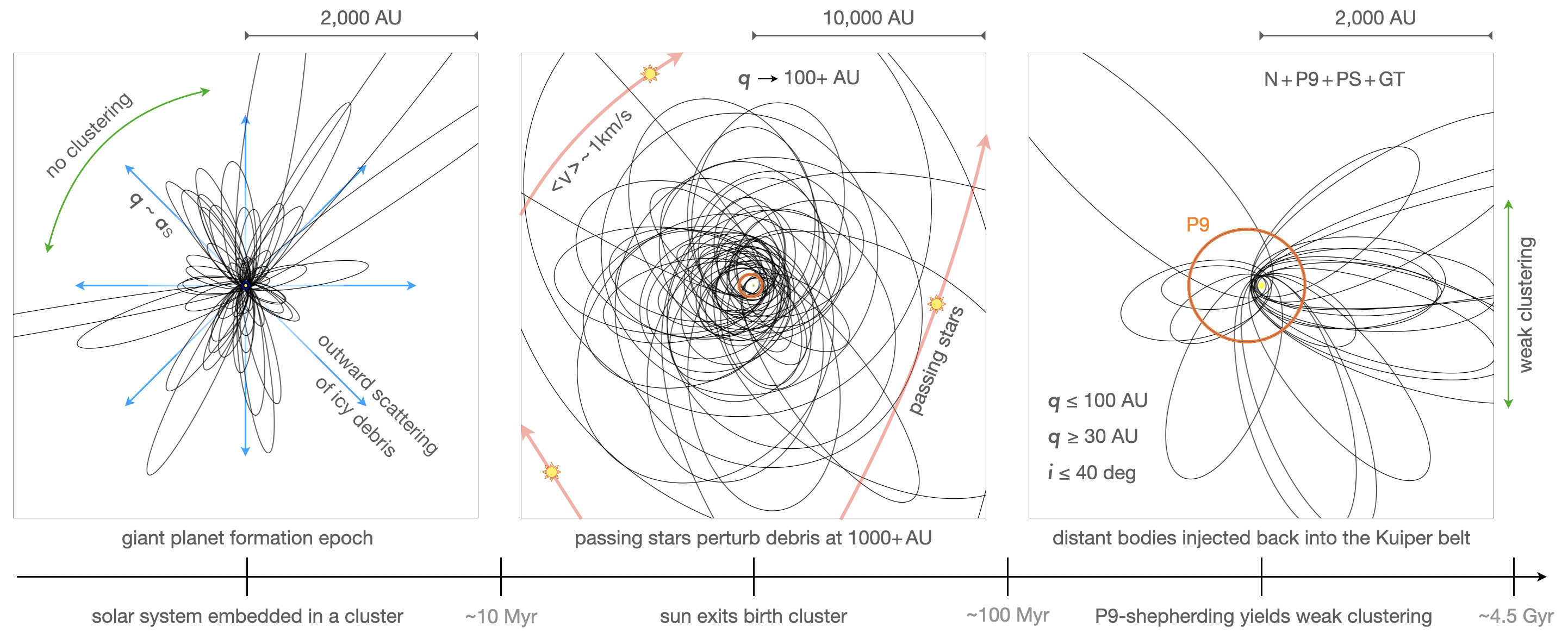}
\caption{Sequence of events modeled within this work. A perihelion-detached population of trans-Neptunian objects forms while the sun is embedded within its birth cluster. Subsequently, over the multi-Gyr lifetime of the solar system, Planet Nine slowly diminishes the perihelia of a subset of these extremely long-period objects, mixing them into the observed census of Kuiper belt objects. Orbits shown on the left and middle panels of the Figure are taken from the $t=4\,$Myr model of the inner Oort cloud (IOC) shown in Figure (\ref{F:cluster}). Orbits depicted on the right panel of the Figure correspond to the same sample of orbits evolved for $t>2\,$Gyr, and shown as dynamical footprints Figure (\ref{F:sequence}).
}
\label{F:sequence}
\end{figure*} 

Although the aforementioned picture reasonably represents the evolution of objects with semi-major axes on the order of a few hundred AU, more recent detections of trans-Neptunian objects \citep{2016AJ....152..221S,2019AJ....157..139S} increasingly point to a pronounced abundance of long-period TNOs with $a\gtrsim1000\,$AU. Crucially, this orbital domain borders the inner Oort cloud (IOC) -- a population of debris captured onto detached, long-term stable orbits during the solar system’s residence within its birth cluster \citep{2011Icar..215..491K,2020AJ....159..285C}. The marked existence of such extremely long-period members of the Kuiper belt demands that we consider the possibility that a fraction of the objects that comprise the population shown in Figure (\ref{F:orbits}) have been injected into the distant solar system from the outside, through an intricate interplay between Galactic effects and Planet Nine’s gravity. 

In this work, we investigate this hypothesis through numerical experimentation, with an eye towards understanding how the process of mixing between the outer Kuiper belt and the IOC alters the degree of orbital confinement facilitated by Planet Nine. In particular, we simulate the formation of the inner Oort cloud within the solar system’s birth cluster (section \ref{sec:FIOC}) and subject the resulting population of debris to P9-forced evolution (section \ref{sec:P9E}). As we describe below, our results indicate that although material injected into the distant Kuiper belt by Planet Nine exhibits orbital clustering, the degree of confinement is inferior compared to that exhibited by objects that originate within the scattered disk. These findings introduce an added degree of uncertainty into our attempts to accurately estimate the physical and orbital properties of Planet Nine.

\section{Formation of the Inner Oort Cloud} \label{sec:FIOC}

Statistically speaking, the vast majority of stars form in stellar associations \citep{2003ARA&A..41...57L,2003AJ....126.1916P}. The solar system is no exception to this rule, and the enrichment of meteorites in short-lived radiogenic isotopes is among the most direct lines of evidence that point to the notion that the sun's birth environment likely played an important role in shaping the solar system's large-scale structure (see \citealt{2010ARA&A..48...47A} for a review). In fact, if the process of planetary accretion unfolded while the sun was still embedded within its birth cluster, the formation of a perihelion-detached cloud of debris would have been all but inevitable. 

An inescapable consequence of the solar system's residence within a cluster of stars is that at sufficiently large orbital radii, passing stars and the cluster's mean-field potential exert significant perturbations on solar system objects. Correspondingly, any planetesimals that would have been scattered out to large heliocentric distances by the forming giant planets would have experienced some degree of modulation by cluster effects \citep{2004AJ....128.2564M}. In particular, published numerical simulations show that as long as the solar system continues to reside within its birth cluster, the population of planetesimals that attain $r\gtrsim1000\,$AU through outward scattering remain entrained in a quasi-steady state cycle of enrichment through giant planet scattering, circularization through cluster effects, and stripping by passing stars \citep{Brasser2006,2012Icar..217....1B}. As soon as the sun leaves its birth environment, however, this process ceases to operate, terminating the generation of the IOC.

We have simulated the sequence of events depicted in Figure (\ref{F:sequence}), following the procedure outlined in \citet{Brasser2006}. Our simulations included Jupiter and Saturn, residing at $a_{\rm{J}}=5.5\,$AU and $a_{\rm{S}}=9\,$AU respectively, as well as a sea of $N=10^5$ massless planetesimals, spanning the $4.5-12\,$AU range in heliocentric distance. Both of the planets as well as the test particles were initialized on circular and coplanar orbits, and were integrated using the conservative variant of the Bulisch-Stoer algorithm implemented in the \texttt{mercury6} gravitational dynamics software package \citep{1999MNRAS.304..793C}. The initial time step and integration accuracy parameter were set to $\Delta t=10^2$ days and $\epsilon=10^{-12}$ respectively.

\begin{figure*}[tbp]
\centering
\includegraphics[width=\textwidth]{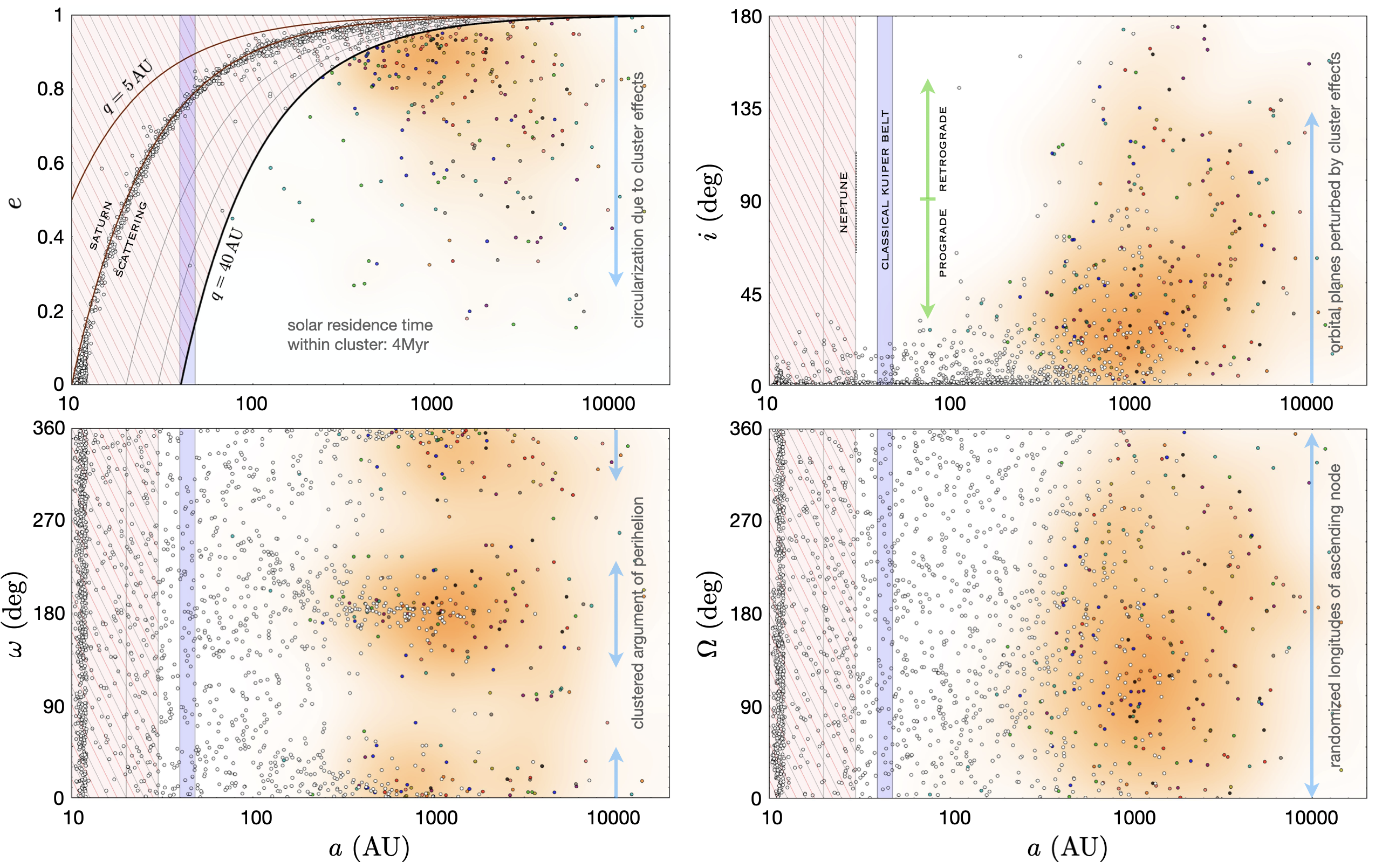}
\caption{A representative inner Oort cloud created in our simulations. Multi-colored points depict perihelion-detached population of TNOs, generated within ten distinct realizations of the model cluster. Crudely speaking, cluster effects facilitate perihelion lifting, excitation of inclination, and clustering of the argument of perihelion for $a\gtrsim500\,$AU objects. As the sun's cluster residence time increases, both the inner and outer edges of the detached population inch towards smaller heliocentric distances.}
\label{F:cluster}
\end{figure*} 

In addition to planetary perturbations, we accounted for the effects of passing stars with impact parameters smaller than $b_{\star}\leqslant0.1\,$pc, as well as the mean-field potential of the solar birth cluster. For definitiveness, we modeled the cluster as a Plummer sphere, setting the total mass to $M_{\infty}=1200\,M_{\odot}$ and Plummer radius to $c=0.35\,$pc in an effort to approximately mimic the properties of the Orion Nebular Cluster\footnote{Note that these choices correspond to a cluster core radius of $r_{\rm{c}} \approx 0.23\,$pc and central number density of $n_{\rm{c}}\approx 1.7\times10^4/$pc$^3$ -- values comparable to the properties of the Trapezium cluster, embedded within the ONC. Moreover, with these parameters, the mean stellar number density interior to the $M/M_{\infty} = 95\%$ radius evaluates to $\langle n \rangle\approx100/$pc$^3$, in agreement with observations \citep{1998ApJ...492..540H}.} (see also \citealt{2020AJ....159..101B} and the references therein). We further assumed the sun to steadily orbit the cluster core at the half-mass radius of $r_{\odot}=(1+2^{1/3})/\sqrt{3}\,c\approx 0.33\,$pc on a circular orbit, and ignored the long-term dissipation of the system, thereby holding the stellar number density fixed at $n_{\odot}\approx1400/$pc$^3$. Finally, we adopted a velocity dispersion of $\langle v \rangle=1\,$km/s for all 19 stellar species\footnote{White dwarfs were neglected from the stellar mass function.} that comprised the model stellar mass function of \citet{HTA87}, and intermittently injected passing stars into the simulation domain following the method outlined in \citet{HTA87}. Unlike these authors, however, we did not employ the impulse approximation, and instead resolved stellar flybys through self-consistent $N$-body integration (adopting the simulation parameters described above), setting the stellar injection/ejection radius to $0.1\,$pc. Cumulatively, these choices yield a characteristic stellar encounter rate of $\Gamma\sim n_{\odot}\,\pi\,b_{\star}^2\,\langle v \rangle\sim50/$Myr.

For completeness, we note a number of additional simplifying assumptions that were made in our calculations. First, we assumed an essentially perfect star-formation efficiency, such that that the total mass of cluster is equal to the mass comprised by the stars. In practice, this correspondence is of little dynamical consequence, since Kozai-Lidov type variations in particle parameters due to the cluster's mean-field potential (see e.g., section 2 of \citealt{2020AJ....159..101B}) represent only a small, longer-term correction to direct perturbations exerted by the stars, meaning that modulating the mass budget of the gas is unlikely to appreciably change the simulations outcomes. Second, we approximated the IMF within the cluster as being similar to the present-day mass function with the exception of omitting white dwarfs. While a more thorough treatment of the IMF is possible, this assumption is unlikely to manifest as the dominant source of error in our calculations. Finally, we ignored the potential presence of P9 itself -- and its accompanying perturbations -- during the cluster stage. This simplification can be justified through a timescale argument: published simulations indicate that P9-induced evolution unfolds on $\sim$Gyr timescales -- far longer than lifetimes of a typical birth cluster. This means that even if Planet Nine were extant during the sun's cluster phase, its gravity would not constitute an efficient perihelion-detachment mechanism for outward scattered planetesimals.

\begin{figure*}[tbp]
\centering
\includegraphics[width=\textwidth]{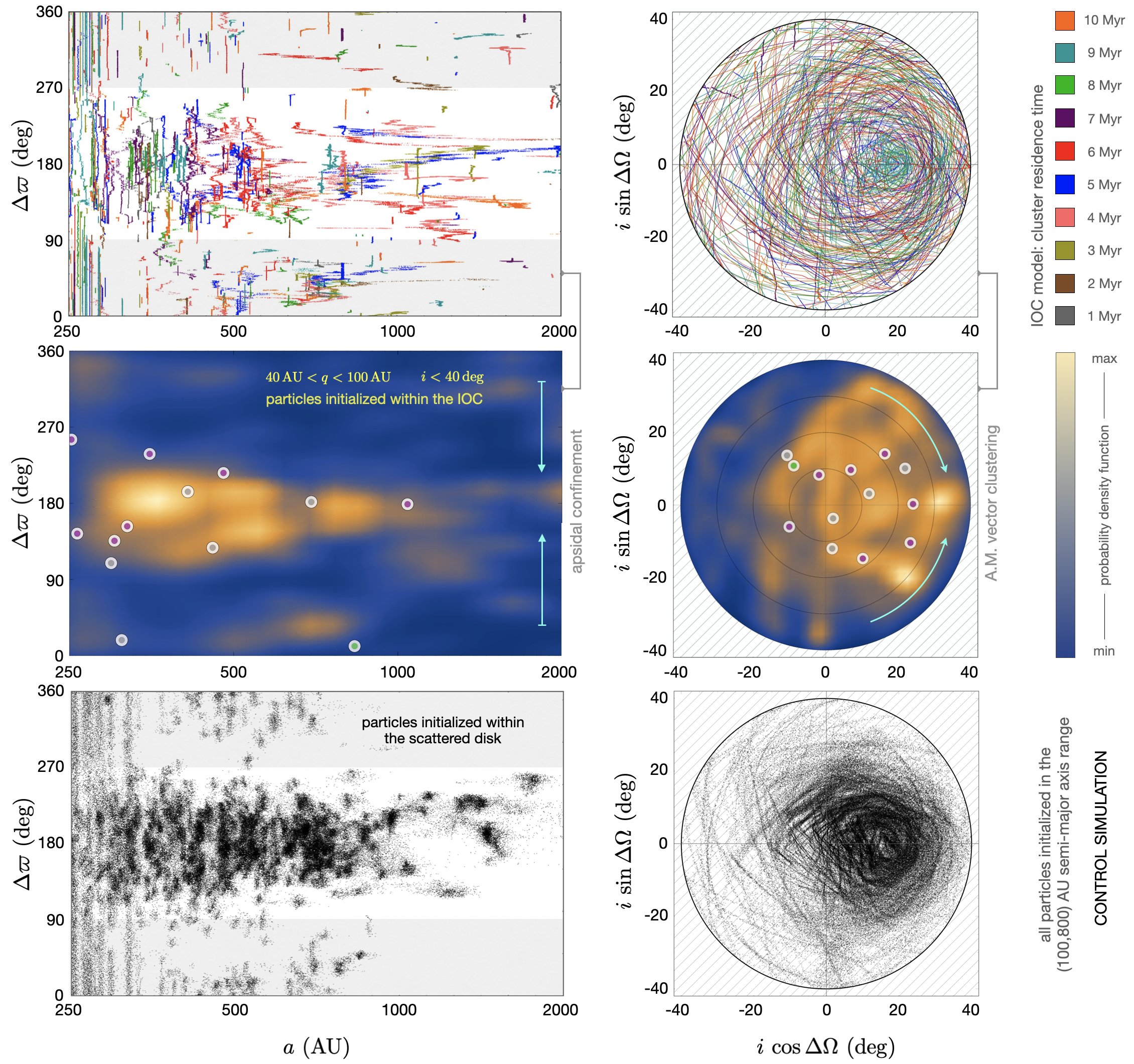}
\caption{Synthetic long-period Kuiper belt. The top panels show the $t>2\,$Gyr, $40\,\mathrm{AU}<q<100\,\mathrm{AU}$, $i<40\,\deg$ dynamical footprints of particles generated solely by diffusion of inner Oort cloud objects to smaller orbital radii. Each set of trajectories is color coded by the cluster residence time of the inner Oort cloud model from which they were generated. The middle panels depict the smoothed density histogram produced by combining all of the results. The depicted data points correspond to the $q>40\,$AU subset of orbits illustrated in Figure (\ref{F:orbits}), and are shown using the same color scheme. Notably, with the exception of a single object (2015 KG$_{163}$), the depicted data correspond to TNOs that do \textit{not} experience rapid orbital diffusion induced by Neptune. The bottom panels show equivalent orbital footprints from a control simulation employing the same Planet Nine parameters, but with all particles initialized uniformly within the $a\in(100,800)\,$AU range. Cumulatively, our results indicate that although P9-facilitated apsidal alignement and angular momentum vector clustering is clearly evident in the IOC-fed simulation results, it is considerably weaker than orbital confinement exhibited by particles sourced from the scattered disk population of Kuiper belt.}
\label{F:results}
\end{figure*} 

To smooth over statistical fluctuations inherent to each recapitulation of stochastic evolution facilitated by stellar encounters, we split the test particle population of our model into ten separate disks (each composed of $N=10^4$ particles), and integrated each system within a distinct realization of the cluster. Each integration was carried out over a timespan of $\tau=10\,$Myr. This timescale is simultaneously comparable to the characteristic lifetime of a typical protoplanetary disk \citep{Haisch2001}, and close to the upper bound on the sun's permissible residence time within the model cluster. Importantly, the latter constraint comes from a dynamical limit on the product of stellar number density and cluster residence time of $\langle n \rangle\,\tau\lesssim 2\times10^4\,$Myr/pc$^3$, derived from the inclination dispersion of the cold classical Kuiper belt \citep{2020AJ....159..101B}, and is largely independent of the processes modeled here. With the adopted parameters, we found that a substantial IOC forms within $\sim1\,$Myr, and reaches steady-state in terms of particle count at $\sim2\,$Myr. More specifically, between $2$ and $10\,$Myr, the number of particles with semi-major axes in excess of $40\,$AU and perihelion distance greater than $40\,$AU fluctuates between 200 and 300, although it is important to understand that the particles that comprise the cloud are continuously in flux. 

A representative snapshot of the dynamical state of the IOC at $t=4\,$My is shown in Figure (\ref{F:cluster}). The top left panel depicts particles eccentricities as a function of semi-major axes. Constant-perihelion curves corresponding to $q=5,10,20,30,$ and $40\,$AU are also outlined, and the outward transport of particles along the Saturnian scattered disk is evident. Objects with $q\leqslant40\,$AU are shown as white circles, while perturbed particles with $q>40\,$AU are denoted with filled points, each color-coded in accord with the cluster realization within which it was generated. A smoothed orange density histogram underlying the perihelion-detached population of particles is also illustrated in each panel. Particle inclinations are depicted on the top right panel of Figure (\ref{F:cluster}) and show that although stellar encounters can generate retrograde objects, the kernel of the IOC is comprised of objects with inclinations on the order of a few tens of degrees. 

The argument of perihelion and longitude of ascending node are portrayed in the bottom left and bottom right panels respectively. While the nodal structure of the cloud is fully randomized, the argument of perihelion exhibits distinct clustering around $\omega=0$ and $\omega=180\,\deg$. Unlike the dynamical grouping shown in Figure (\ref{F:orbits}), these patterns are readily understood. Recall that within the context of our simulations, particles attain large heliocentric distances exclusively by scattering off of giant planets, meaning that by the time they get perturbed by passing stars, their orbits are nearly radial but planar, with $e\sim1- q_{\rm{S}}/a\sim0.99$ and $i\sim0$. Because $\Omega$ is ill-defined at $i=0$, even a weak stellar flyby can drastically alter the longitude of ascending node. Significantly changing $\omega$ on the other hand, would require a perturbation to rotate the nearly-radial orbit away from the orbital plane of Jupiter and Saturn by a large angle. Instead, Kepler's second law insinuates that interactions are most likely to occur at aphelion, where an impulsive change in particle velocity can easily modulate the semi-major axis and eccentricity, but not alter the orbital orientation. As a result, even after being perturbed by passing stars, particles continue to come to aphelion close to the orbital plane of the giant planets, thereby maintaining $\omega=0$ or $\omega=180\deg$. 

The dynamical state of the IOC appears similar at other times in the integration, with the caveat that the inner and outer boundaries of the perturbed objects steadily march towards lower heliocentric distances in time, crudely tracking the ever-decreasing impact parameter of the closest stellar encounter $b_{\rm{min}}\sim1/\sqrt{\pi\,n_{\odot}\,\langle v \rangle\,t}$. Overall, we find that the dynamical decoherence timescale of the detached population is somewhat shorter than a million years, and adopt snapshots of the distant solar system at $1,2,3,...,10\,$Myr as pseudo-independent models of the IOC. 

\section{P9-Induced Evolution} \label{sec:P9E}

\begin{figure*}[tbph]
\includegraphics[width=\textwidth]{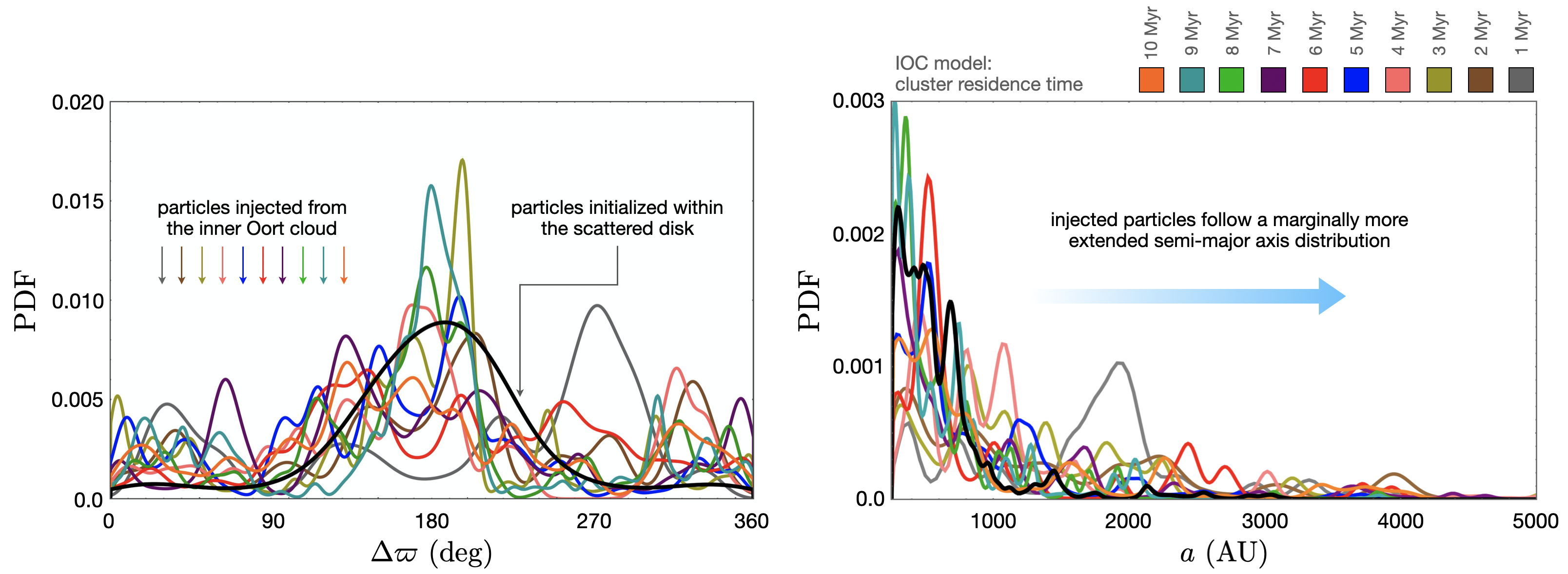}
\caption{Apsidal clustering and semi-major axis distributions of simulated particles satisfying $a>250\,$AU, $40\,\mathrm{AU}<q<100\,\mathrm{AU}$, $i<40\,\deg$ and $t>2\,$Gyr. The multi-colored curves show the $\Delta\varpi$ and $a$ probability density functions of inward-injected Oort cloud objects obtained in our simulations. The black curves depict the probability density functions associated with synthetic KBOs that are initialized within the scattered disk, and evolved with identical ($m_9=5\,M_{\oplus}, a_9=500\,$AU, $e_9=0.25$, $i_9=20\,\deg$) P9 parameters. Although the dynamical forcing is the same, objects that originate within the Kuiper belt exhibit significantly tighter apsidal clustering in our simulations than do injected IOC objects. Injected IOC objects further display a somewhat more extended semi-major axis distribution than our control simulation.}
\label{F:apsidal}
\end{figure*} 

In absence of Planet Nine, the cloud of distant debris created by the interplay between giant planet scattering and cluster perturbations would essentially remain frozen over the main-sequence lifetime of the sun. Gravitational perturbations due to P9, however, can steadily diminish the perihelia of such detached objects, bringing them into the fold of the scattered disk population of the Kuiper belt. Here, we have simulated this process, self-consistently accounting for the $N-$body dynamics driven by Neptune, Planet Nine, passing stars within the Galactic field (once again, following the methodology of \citealt{HTA87}), as well as the effect of the Galactic tide (implemented as in \citealt{2017ApJ...845...27N}; see also \citealt{HT86,1999Icar..137...84W,2020AJ....159..285C}). As noted in the aforementioned references, this effect is most sensitive to the vertical component of the acceleration, which is largely determined by the mean local Milky Way disk density, which we set to $\rho_{\rm{MW}}=0.1\,M_{\odot}/\rm{pc}^3$. 

For definitiveness, we adopted the best-fit P9 parameters from \citep{BABB}: $m_9=5\,M_{\oplus}, a_9=500\,$AU, $e_9=0.25$ and $i_9=20\,\deg$. Orbit-averaged effects of Jupiter, Saturn, and Uranus were also included in the calculations as an effective $J_2$ moment of the sun. As discussed in \citet{BABB}, this numerical scheme constitutes an adequate compromise between computational cost and accuracy, allowing us to increase the initial tilmestep in our simulations to $\Delta t=3000\,$days.

We integrated each of the ten variants of the IOC discussed in the previous section for $4.5\,$Gyr, and found that over the lifetime of the sun, a significant fraction (that is, on the order of $20\%$) of the IOC gets injected into the distant ($a>250\,$AU) Kuiper belt, with approximately $10\%$ injected in the final $2\,$Gyr. Dynamical footprints with $t>2\,$Gyr of injected particles that satisfy crude stability and observability criteria ($40\,\mathrm{AU}<q<100\,\mathrm{AU}$, $i<40\,\deg$) are shown in the top panels of Figure (\ref{F:results}), where the color of the points informs the IOC model adopted for the initial conditions. More specifically, the panels on the left hand side depict the particles' longitude of perihelion, relative to that of Planet Nine, while the polar plots on the right inform the angular momentum vectors of particle orbits.

Overall, IOC models that correspond to solar cluster residence time of $t\geqslant2\,$Myr yield consistent results, although the relevant particle count in each case is relatively small (Table \ref{frequencies}). Thus, in order to more clearly portray the orbital distribution of the long-period Kuiper belt generated by inward-injection of distant debris, we show a smooth density histogram of our combined\footnote{Although the distant Kuiper belt sourced from the $t=1\,$Myr model of the IOC shows virtually no apsidally anti-aligned objects with $q<100\,$AU, the particle count is a factor of a few smaller than the other runs, meaning that its contribution is insignificant.} simulation data in the middle panels of Figure (\ref{F:results}). The census of known long-period KBOs satisfying the same perihelion and inclination cuts is over-plotted on the digram, using the same color scheme as that in Figure (\ref{F:orbits}). Finally, the bottom panels of Figure (\ref{F:results}) depict the results of a control simulation using the same P9 parameters, but with particle initial conditions drawn uniformly from the $q\in(30,100)$\,AU and $a\in(100,800)\,$AU ranges, and an initial inclination dispersion corresponding to a half-gaussian distribution with a standard deviation of $15\deg$.

\begin{table*}[t]
\centering
\caption{A summary of orbital confinement metrics for each IOC-fed simulation. The percentages in parentheses indicate the fraction of down-sampled control simulations that yield metrics as low or lower than the reported numerical experiments.}
\begin{tabular}{l l l l l l l l l l l l }
\hline\hline
IOC model & 1 & 2 & 3 & 4 & 5  \\ 
\hline
$N^{\dagger}$ & 11 & 15 & 15 & 26 & 37  \\
$f_\varpi$                                   & 0.46 (0\%)     & 0.62 (0\%)   & 0.69 (0\%)  & 0.64 (0\%) & 0.75 (0\%)    \\
$\langle i \rangle$ (deg)             & 20.2 (100\%) & 9.2 (40\%)   & 10.1 (67\%) & 4.6. (0\%) &  4.2  (0\%)    \\
$\langle i \rangle/i_{\rm{rms}}$ & 0.98 (100\%)   & 0.37 (23\%) & 0.43 (26\%) & 0.20 (0\%) & 0.16 (0\%)  \\
\hline\hline
IOC model & 6 & 7 & 8 & 9 & 10  \\ 
\hline
$N^{\dagger}$ & 31 & 28 & 30 & 30 & 37  \\
$f_\varpi$                                    & 0.70 (0\%) & 0.65 (0\%)  & 0.71 (0\%)  & 0.64 (0\%)  & 0.67 (0\%)   \\
$\langle i \rangle$ (deg)             & 7.4  (10\%)& 0.5 (0\%)    & 3.8  (0\%)  & 10.7 (97\%) & 3.6 (0\%)  \\
$\langle i \rangle/i_{\rm{rms}}$ &  0.29 (0\%)&  0.02 (0\%)  & 0.14 (0\%)  & 0.45 (47\%) & 0.13 (0\%)  \\
\hline
\end{tabular}
\label{frequencies}
\end{table*}

Cumulatively, our simulations elicit two key results. First, the injected particles clearly follow typical P9-driven dynamical evolution (see \citealt{BB2016,2017AJ....154..229B,2018AJ....155..250K, 2020PASP..132l4401K,2018AJ....156..263L,2018AJ....155..249H}) with relative longitude of perihelion exhibiting visible clustering around $\Delta\varpi\sim180\deg$ (Figure \ref{F:results}). Similarly, the inclination - node trajectories of these objects circulate around the P9-forced equilibrium in phase space, which can manifest in an apparent clustering of the longitudes of ascending node when viewed from the ecliptic. At the same time, it is important to note that the patterns of orbital clustering shown in Figure (\ref{F:results}) are considerably less pronounced than those created in P9 simulations where distant KBOs are sourced entirely from the scattered disk.

To be more specific, the fraction of dynamical footprints that fall within $\pm90\deg$ of exact apsidal anti-alignment with respect to P9 in the middle panel of Figure (\ref{F:results}) is a mere $f_{\varpi}=67\%$. By comparison, our control simulation with initial semi-major axes of particles drawn from the $a\in(100\mathrm{AU},800\mathrm{AU})$ range yields $f_{\varpi}=88\%$. The confinement of angular momentum vectors in our IOC-fed simulations is similarly weak: the mean inclination and the associated rms dispersion exhibited by injected particles are $\langle i \rangle\approx5\,\deg$ and $i_{\rm{rms}} \approx26\,\deg$, respectively. Meanwhile, our scattered disk-fed control simulation shows a more tightly clustered particle distribution, characterized by $\langle i \rangle\approx9\,\deg$ and $i_{\rm{rms}}\approx19\,\deg$. Thus, taken at face value, our calculations suggest that the presence of a significant IOC could partially obfuscate the dynamical signal generated by Planet Nine.

The difference in the statistical measures of orbital clustering between inward-injected IOC objects and outward-scattered Kuiper belt objects within the context of P9 simulations begs the question of whether or not this discrepancy can be attributed to a difference in particle count among the various numerical experiments. A cursory examination of our simulation suite, however, suggests that this is unlikely to be the case. Out of the 2440 objects initialized within the full set of our IOC-fed simulations, a total of 260 particles attain orbits that satisfy the aforementioned observability criteria at some time in excess of $t>2\,$Gyr. The same is true for 227 (out of 1000 initial) particles in our control simulation, implying that the dynamical footprints depicted in the top and bottom panels of Figure (\ref{F:results}) are derived from a comparable number of independent tracers.

In order to quantify the statistical significance of these differences further, we computed the three measures of dynamical confinement -- $f_{\varpi}$, $\langle i \rangle$, $\langle i \rangle/i_{\rm{rms}}$ -- for each IOC-fed model independently, and computed the likelihood that values as low as these can arise within the context for an appropriately re-sampled control simulation, where the particles are taken to originate within the Kuiper belt. In particular, we down-sampled the control simulation such that the number of independent particles that satisfy our observability criteria at $t > 2\,$Gyr, $N^{\dagger}$, are the same as that in each IOC-fed simulation, and computed the orbital confinement metrics thirty times, randomly reshuffling the particles. We then recorded the fraction of such reshuffled simulations that yield values of $f_{\varpi}$, $\langle i \rangle$, and $\langle i \rangle/i_{\rm{rms}}$ equal to or lower than those obtained within the corresponding IOC-fed calculation.

These results are summarized in Table \ref{frequencies}. As can be deduced from these calculations, it is improbable that the differences in the synthetic Kuiper belts sourced from inward injection vs. outward scattering are spurious. In particular, although the inclination-node clustering generated in IOC-fed calculations are in some cases consistent with our control simulation, apsidal clustering in these simulations is consistently inferior. To illustrate this point further, we show the $\Delta\varpi$ distributions of simulated particles in the left panel of Figure (\ref{F:apsidal}), where the control and IOC-fed simulations are shown with black and multi-colored curves, respectively.

A second important result of our numerical experiments is that steady infusion of IOC material into the Kuiper belt can enhance the prevalence of scattered disk objects with semi-major axes in excess of $a\gtrsim1,000\,$AU. That is, although the semi-major axis distribution of dynamical footprints shown in Figure (\ref{F:results}) peaks in the $a\sim250-500\,$AU range, the tail of the distribution extends beyond $a\gtrsim3000\,$AU, as shown in the right panel of Figure (\ref{F:apsidal}). Relative to the results of our control simulation, these IOC-fed calculations can readily yield a factor of $\sim5$ enhancement in the semi-major axis distribution of distant KBOs at $a=2000\,$AU and more than an order of magnitude increase at $a=3000\,$AU. Thus, an extended dispersion of KBO semi-major axes may indirectly hint at the operation of P9-facilitated modulation of the IOC.

Simultaneously, however, it must be understood that this determination is preliminary. Recall that while the initial conditions of our IOC-fed calculations are generated through self-consistent numerical experiments, the initial conditions of our control simulation were chosen in an \textit{ad-hoc} manner. Because the starting semi-major axis distribution of our control calculation is truncated at 800 AU, the results may be underestimating the breadth of the true semi-major axis distribution that can be generated through P9 induced evolution of the scattered disk. On the other hand, if the intrinsic surface density of the scattered disk diminishes more steeply than $\Sigma_{\rm{SD}}\propto1/r$ (as is suggested for example by the results of \citealt{2017ApJ...845...27N}), then our control simulation is likely to be overestimating the extent of the high-$a$ tail of the orbital distribution. A distinct degeneracy relevant to the surface density profile of distant icy bodies arises from the possibility that the presence of the classical Oort cloud may generate an uptick in TNOs with $a\gtrsim1{,}000\,$AU that is completely independent of P9-facilitated evolution. Therefore, more precise calculations that account for the existence of the classical Oort cloud and employ a self-consistent estimate of the initial conditions of the scattered disk, are necessary to confidently elucidate the difference between the semi-major axis distributions formed by IOC-fed and scattered disk-sourced populations of distant TNOs.

\section{Conclusion}

In this work, we have carried out a series of numerical experiments in an effort to quantify the injection of inner Oort cloud material into the distant Kuiper belt via P9-induced dynamical evolution. To this end, we have constructed a model IOC, accounting for the gravitational effects of the sun’s birth cluster (section \ref{sec:FIOC}), and have followed the long-term evolution of this population of distant debris, subject to perturbations from the giant planets, passing stars, galactic tide, as well as P9 (section \ref{sec:P9E}). The results of our simulations are readily summarized: if Planet Nine exists, then it facilitates steady variations in the orbits of $a\gtrsim1000\,$AU IOC bodies, bringing a fraction of them into the fold of the observable component of the distant scattered disk. Much like the conventional members of the long-period Kuiper belt, these injected objects exhibit P9-driven orbital confinement. However, the degree of clustering is considerably weaker.

This discrepancy carries important observational consequences. Because the extent of apsidal clustering of distant objects \textit{increases} with $e_9$ \citep{BABB}, contamination of the distant Kuiper belt by poorly-confined IOC objects implies that a more eccentric and distant Planet Nine may be required to explain the data, than is insinuated by the existing literature. However, the magnitude to which a best-fit P9 orbit must be modified depends on a yet-unconstrained fraction of KBOs that are sourced from the IOC. Irrespective of this notion, it is important to note that Planet Nine's eccentricity cannot be enhanced by a large margin without violating other constraints, including the over-production of highly inclined scattering objects that ensues for sufficiently high values of  $e_9$, as pointed out by \citet{2019AJ....158...43K}. Cumulatively, the proof-of-principle results reported herein introduce an important, additional degree of uncertainty in our efforts to pin-down the orbit of Planet Nine.
\\
\begin{acknowledgments}
We are thankful to Alessandro Morbidelli, Darryl Seligman, Juliette Becker, Fred Adams, David Nesvorn{\'y}, and Eduardo Marturet for insightful discussions. We are additionally grateful to David Nesvorn{\'y} for sharing his implementation of Galactic tide accelerations. We thank the anonymous referee for  providing a thorough and insightful report. K.B. is grateful to Caltech, the David and Lucile Packard Foundation, and the Alfred P. Sloan Foundation for their generous support. 
\end{acknowledgments}

\end{document}